\documentclass[aps,prd,reprint,nofootinbib,superscriptaddress,floatfix]{revtex4-2}
\usepackage[T1]{fontenc}
\usepackage{lmodern}
\usepackage{microtype}
\usepackage{amsmath,amssymb,amsthm,mathtools,bm}
\usepackage{booktabs}
\usepackage{graphicx}
\usepackage{enumitem}
\usepackage{xcolor}
\usepackage[colorlinks=true,linkcolor=blue!45!black,citecolor=blue!45!black,urlcolor=blue!45!black]{hyperref}
\allowdisplaybreaks
\newtheorem{theorem}{Theorem}[section]
\newtheorem{proposition}[theorem]{Proposition}
\newtheorem{lemma}[theorem]{Lemma}
\newtheorem{corollary}[theorem]{Corollary}

\newtheorem{remark}[theorem]{Remark}

\newcommand{\RR}{\mathbb R}

\newcommand{\cD}{\mathcal D}

\newcommand{\cT}{\mathcal T}
\newcommand{\one}{\mathbf 1}
\newcommand{\norm}[1]{\left\lVert #1\right\rVert}
\newcommand{\abs}[1]{\left| #1\right|}
\newcommand{\dd}{\mathrm d}
\newcommand{\vol}{\operatorname{vol}}

\hypersetup{
  pdftitle={Optical and ADM Tomography from Directed Causal Arrival Times},
  pdfauthor={Maxim V. Churilov}
}

\makeatletter
\AtBeginDocument{%
  \@ifpackageloaded{hyperref}{\hypersetup{hidelinks}}{}%
}
\makeatother

\begin{document}

\title{Optical and ADM Tomography from Directed Causal Arrival Times}
\author{Maxim V. Churilov}
\email{churilovm1305@gmail.com}
\affiliation{Independent Researcher, Orenburg, Russia}
\date{October 26, 2026}

\begin{abstract}
We formulate a local reconstruction of Lorentzian geometry from directed null arrival times and spacetime volume, assuming a supplied clock foliation and spatial coordinate gauge. Static arrival data recover the optical metric $\bar h=N^{-2}h$, while a normalized volume density fixes the remaining conformal scale up to a global constant. In the stationary case, the Randers arrival function separates optical geometry and shift. For time-dependent ADM metrics we derive the exact future null indicatrix, its algebraic inverse, and a minimal finite-direction sampling scheme: $d(d+3)/2$ one-way arrival measurements suffice in $d$ spatial dimensions, are necessary for stable local recovery within the specified scalar-measurement class, and admit an exact semialgebraic consistency test with certified robustness and rejection margins under bounded measurement error. A short-time freezing estimate and finite-resolution inverse give metric error $O(a/\epsilon+\epsilon+\zeta)$ for arrival error $a$, probe scale $\epsilon$, and volume error $\zeta$; optimizing yields $O(\sqrt a+\zeta)$. Algebraic and numerical checks reproduce the inversion and stability formulas. The result is an operational, gauge-conditional tomography theorem; it does not infer the clocks or arrival oracle from a background-free quantum theory.
\end{abstract}

\maketitle

\section{Introduction}
\label{sec:intro}

The classical causal-reconstruction theorems establish a powerful but deliberately qualitative statement: under appropriate causality assumptions, chronology determines the topology, differentiable structure, and conformal class of a spacetime metric \cite{HawkingKingMcCarthy1976,Malament1977,Minguzzi2019}.  A volume form then fixes the conformal factor.  This is the continuum origin of the causal-set slogan ``order plus number equals geometry,'' and it motivates ordered-measure formulations of finite spacetime data \cite{BombelliNoldusTafoya2014}.  Yet two operational questions remain separate from that theorem.

First, what is actually reconstructed when an experiment measures a signal arrival time between two spatially labelled laboratories?  In a static but non-ultrastatic spacetime, the answer is the optical metric, not the instantaneous spatial metric.  In a stationary spacetime with drift or rotation, the answer is a nonreversible Randers geometry.  The stationary-to-Randers correspondence and relativistic Fermat principle are classical \cite{CaponioJavaloyesMasiello2011,Javaloyes2012,CaponioJavaloyes2026}; the ADM extension to a time-parameterized Randers structure is likewise known at the level of causal encoding \cite{SkakalaVisser2011}.  What is missing for finite operational reconstruction is an explicit inverse from a small set of directed arrival measurements, together with stability and scale selection.

Second, a global inverse problem and a local calibration problem are not the same.  Boundary rigidity and light-observation methods recover Lorentzian or Finsler geometry from nonlocal data under geometric hypotheses \cite{KurylevLassasUhlmann2018,HintzUhlmann2019,UhlmannYangZhou2021,Monkkonen2022,deHoopIlmavirtaKykkanenSaksala2026}.  Here the data are different: a clock-labelled observer can launch a signal through an infinitesimal displacement in selected local directions and record the first threshold crossing.  The target is the instantaneous null cone and its ADM coefficients at the emission event.  This local problem admits a finite algebraic solution even when the spacetime is not stationary.

The present paper develops that solution.  Its central chain is
\begin{equation}
 \text{directed arrivals}
 \longrightarrow F_z
 \longrightarrow (a_z,\omega_z)
 \longrightarrow [g_z]
 \xrightarrow{\text{volume}} g_z,
 \label{eq:chain}
\end{equation}
where $[g_z]$ denotes the conformal class.  The first arrow is dynamical and finite-resolution; the remaining arrows are local algebra.

The main results are as follows.
\begin{enumerate}[label=(\roman*)]
\item For static metrics $g=-N^2\dd t^2+h$, the chronological relation is exactly the optical-distance inequality $d_{N^{-2}h}(x,y)<t-s$.  Order and normalized volume recover $N$ and $h$ up to a single constant conformal factor.
\item For standard stationary metrics $g=-N^2(\dd t-\vartheta)^2+h$, directed arrival times form the Randers distance of $F=\sqrt{N^{-2}h}+\vartheta$.  The even and odd parts of $F$ recover the optical tensor and the one-form $\vartheta$ separately.
\item For a general ADM metric, the instantaneous arrival indicatrix is derived exactly and inverted in closed form.  The cone determines $g$ up to a positive factor; the volume density supplies that factor up to one global constant.
\item In $d$ spatial dimensions, a Lorentzian indicatrix in a fixed local frame has
\begin{equation}
 p_d=\frac{d(d+1)}2+d=\frac{d(d+3)}2
 \label{eq:parameter-count}
\end{equation}
independent coefficients.  We exhibit a $p_d$-measurement protocol and prove its $C^1$ local minimality.
\item The local arrival time differs from the frozen indicatrix by $O(\epsilon^2)$.  With absolute oracle error $a$, the normalized error is $a/\epsilon+O(\epsilon)$, so the optimal probe length is $\epsilon\asymp\sqrt a$.
\item A bounded-density anti-concentration lemma converts arrival-time error into a probability that two causal indicators disagree.  This supplies the local interface to finite causal-profile reconstruction without duplicating the global statistical theory of Ref.~\cite{ChurilovFiniteResolution2026}.
\end{enumerate}

The use of a clock foliation is essential and explicit.  A stationary change of section $t\mapsto t+\phi(x)$ sends $\vartheta\mapsto\vartheta+\dd\phi$.  Thus an abstract stationary spacetime determines a gauge class, while a clock-labelled experiment fixes the section and measures a definite representative.  Similarly, the ADM shift depends on the chosen observer congruence.  Our reconstruction is covariant under changes of the supplied coordinates, but it does not construct the coordinates themselves.

Table~\ref{tab:status} records the scope.  The novelty is not the classical fact that null cones encode conformal geometry or that stationary causality is Randersian.  It is the finite directional inverse, the minimal measurement count, the cone--volume formulas in operational variables, and the finite-resolution error law for a genuinely time-dependent ADM metric.

\begin{table*}[t]
\caption{Logical status and scope of the principal statements.}
\label{tab:status}
\begin{ruledtabular}
\small
\begin{tabular}{p{0.26\textwidth}p{0.12\textwidth}p{0.20\textwidth}p{0.31\textwidth}}
Result & Status & Geometric regime & Essential input\\
\hline
Static optical representation & Exact & Static spacetime & Supplied product time $t$\\
Static cone--volume inverse & Exact & Static spacetime & Optical metric and normalized volume density\\
Stationary Randers representation & Classical identity, operationally specialized & Standard stationary spacetime & Clock section and directed arrivals\\
ADM indicatrix inverse & Exact local algebra & General nonstationary ADM metric & Subcritical shift $h(\beta,\beta)<N^2$\\
$p_d$-direction protocol & Exact and locally minimal & Lorentzian/Randers indicatrix & Fixed local spatial frame\\
Short-time freezing law & Exact estimate & Lipschitz time-dependent indicatrix & Uniform convexity and local coordinate control\\
Finite-resolution metric rate & Exact conditional theorem & Local ADM tomography & Arrival oracle, density estimator, calibrated directions\\
Background-free generation of the oracle & Not claimed & Quantum geometry & Independent dynamical theorem required
\end{tabular}
\end{ruledtabular}
\end{table*}

\section{Data model, gauges, and arrival observables}
\label{sec:data}

Let $M$ be a $d$-dimensional smooth manifold and let $I\subset\RR$ be an interval carrying the operational clock coordinate $t$.  A spacetime region is represented as $\mathcal U\subset M\times I$.  We assume throughout that $t$ is a temporal function on $\mathcal U$ and that the spatial labels $x\in M$ are part of the measurement record.

For an emission event $z=(t,x)$ and a nearby spatial displacement $y$ in a coordinate chart, define the first causal arrival time
\begin{equation}
 \cT_z(y)=\inf\{\tau>0:(t,x)\leq (t+\tau,x+y)\},
 \label{eq:arrival-def}
\end{equation}
with the infimum restricted to curves remaining in the chart.  In a stationary spacetime the analogous global quantity depends only on $x,y$ and equals a directed distance.  In a nonstationary spacetime it is a local control value rather than the distance of one fixed Finsler metric.

A finite quantum or classical detector usually provides a thresholded channel time rather than the exact causal boundary.  We therefore allow an oracle $\widehat\cT$ satisfying
\begin{equation}
 \abs{\widehat\cT_z(y)-\cT_z(y)}\leq a
 \label{eq:oracle-error}
\end{equation}
for the requested displacements, possibly outside an exceptional event.  The preceding two papers in this series construct and stabilize such two-sided threshold times in flat and finite-volume lattice models \cite{ChurilovSharpCones2026,ChurilovFiniteVolume2026}.  Here Eq.~\eqref{eq:oracle-error} is an input; the task is geometric inversion.

\begin{remark}[Gauge boundary]
The one-form measured from directional asymmetry is relative to the supplied clock section.  Under $t'=t+\phi(x)$, a standard stationary Fermat one-form transforms by $\vartheta'=\vartheta+\dd\phi$, while the endpoint-labelled directed travel time changes by $\phi(y)-\phi(x)$.  An abstract boundary-distance problem therefore has the familiar potential-field ambiguity \cite{Monkkonen2022}.  In the present local protocol, the clock readings fix $t$, so the representative is observable.
\end{remark}

\section{Static optical reconstruction}
\label{sec:static}

Consider a static metric
\begin{equation}
 g_N=-N(x)^2\dd t^2+h_x,
 \qquad 0<N_-\leq N(x)\leq N_+<\infty,
 \label{eq:static-metric}
\end{equation}
with $M$ compact.  Define the optical metric
\begin{equation}
 \bar h=N^{-2}h.
 \label{eq:optical-metric}
\end{equation}

\begin{theorem}[Static optical chronology]
\label{thm:static-optical}
For $s<t$,
\begin{equation}
 (x,s)\ll_{g_N}(y,t)
 \quad\Longleftrightarrow\quad
 d_{\bar h}(x,y)<t-s.
 \label{eq:static-chronology}
\end{equation}
If an optical minimizer exists, equality gives a future null curve.
\end{theorem}

\begin{proof}
For a future causal curve $\gamma(\lambda)=(x(\lambda),t(\lambda))$,
\begin{equation}
 -N(x)^2\dot t^{\,2}+h_x(\dot x,\dot x)\leq0
 \quad\Longrightarrow\quad
 \dot t\geq\abs{\dot x}_{\bar h}.
 \label{eq:static-causal-speed}
\end{equation}
Integration gives $t-s\geq L_{\bar h}(x(\cdot))\geq d_{\bar h}(x,y)$.  Conversely, if the inequality is strict, choose a spatial curve of optical length below $t-s$ and parameterize its time component with $\dot t>\abs{\dot x}_{\bar h}$.  Equality along an optical minimizer gives a null curve.
\end{proof}

Thus arrival times determine $\bar h$, not $h$.  The missing lapse is encoded in volume.

\begin{theorem}[Static cone--volume reconstruction]
\label{thm:static-volume}
Let the normalized spacetime measure be written relative to the recovered optical metric as
\begin{equation}
 \dd\mu=q(x)\,\dd\vol_{\bar h}(x)\frac{\dd t}{T},
 \qquad q>0.
 \label{eq:static-density}
\end{equation}
Then
\begin{equation}
 q(x)=\frac{N(x)^{d+1}}{\int_M N^{d+1}\dd\vol_{\bar h}}.
 \label{eq:static-q}
\end{equation}
Consequently
\begin{equation}
 N=Cq^{1/(d+1)},\qquad
 h=N^2\bar h,
 \label{eq:static-recovery}
\end{equation}
for one constant $C>0$.  A proper-time calibration or the unnormalized total volume fixes $C$.
\end{theorem}

\begin{proof}
Since $h=N^2\bar h$,
\begin{align}
 \dd\vol_h&=N^d\dd\vol_{\bar h},\notag\\
 \dd\vol_{g_N}&=N\dd\vol_h\dd t
 =N^{d+1}\dd\vol_{\bar h}\dd t.
 \label{eq:static-volume-form}
\end{align}
Normalization gives Eq.~\eqref{eq:static-q}.  The transformation $N\mapsto CN$, $h\mapsto C^2h$ multiplies $g_N$ by $C^2$ but leaves chronology, $\bar h$, and normalized volume unchanged, so the ambiguity is both unavoidable and exhaustive.
\end{proof}

\begin{proposition}[Static stability]
\label{prop:static-stability}
On a class with $q\geq q_->0$, a common fixed global scale
\(C_{\rm sc}\), and uniform ellipticity bounds for \(\bar h\), one has
\begin{equation}
 \norm{N-N'}_{C^0}
 \leq \frac{C_{\rm sc}q_-^{-d/(d+1)}}{d+1}
 \norm{q-q'}_{C^0},
 \label{eq:lapse-stability}
\end{equation}
and
\begin{equation}
 \norm{h-h'}_{C^0}
 \leq C\left(\norm{\bar h-\bar h'}_{C^0}
 +\norm{q-q'}_{C^0}\right).
 \label{eq:static-metric-stability}
\end{equation}
\end{proposition}

\begin{proof}
With the common calibration, Eq.~\eqref{eq:static-recovery} reads
\(N=C_{\rm sc}q^{1/(d+1)}\) and likewise for \(N'\).  The mean-value theorem
on \([q_-,\infty)\) gives Eq.~\eqref{eq:lapse-stability}.  Finally,
\[
 h-h'=N^2(\bar h-\bar h')+(N^2-N'^2)\bar h',
\]
and the uniform bounds on \(N,N'\) and \(\bar h'\), together with the first
estimate, give Eq.~\eqref{eq:static-metric-stability}.
\end{proof}

\section{Stationary directed optical geometry}
\label{sec:stationary}

A standard stationary metric adapted to the supplied clock is
\begin{equation}
 g_{N,\vartheta}=-N(x)^2(\dd t-\vartheta_x)^2+h_x,
 \label{eq:stationary-metric}
\end{equation}
where $\vartheta$ is a one-form and
\begin{equation}
 \bar h=N^{-2}h,
 \qquad \norm{\vartheta}_{\bar h^*}<1.
 \label{eq:stationary-subcritical}
\end{equation}
The associated Fermat function is the Randers norm
\begin{equation}
 F_x(v)=\sqrt{\bar h_x(v,v)}+\vartheta_x(v).
 \label{eq:randers-F}
\end{equation}

\begin{theorem}[Stationary arrival representation]
\label{thm:stationary-arrival}
For $s<t$,
\begin{equation}
 (x,s)\ll_{g_{N,\vartheta}}(y,t)
 \quad\Longleftrightarrow\quad
 d_F(x,y)<t-s,
 \label{eq:stationary-chronology}
\end{equation}
where $d_F$ is the forward distance of Eq.~\eqref{eq:randers-F}.
\end{theorem}

\begin{proof}
For a tangent vector $(\tau,v)$, future causality gives
\begin{equation}
 \tau-\vartheta(v)\geq\frac{\sqrt{h(v,v)}}{N}
 =\sqrt{\bar h(v,v)},
 \label{eq:stationary-null-ineq}
\end{equation}
so $\tau\geq F(v)$.  Integration yields $t-s\geq L_F\geq d_F$.  Conversely, any spatial curve with $F$-length below $t-s$ can be lifted with $\dot t>F(\dot x)$ to a timelike curve.
\end{proof}

The asymmetry $d_F(x,y)\neq d_F(y,x)$ is not a nuisance: it is the observable signature of $\vartheta$.  The complete local split is immediate.

\begin{proposition}[Even--odd Randers split]
\label{prop:even-odd}
For every $v\in T_xM$,
\begin{align}
 \alpha_x(v)&=\frac{F_x(v)+F_x(-v)}2
 =\sqrt{\bar h_x(v,v)},\label{eq:even-odd-alpha}\\
 \vartheta_x(v)&=\frac{F_x(v)-F_x(-v)}2.
 \label{eq:even-odd}
\end{align}
Hence the local directed distance determines $\bar h$ and $\vartheta$ in the chosen clock gauge.
\end{proposition}

The normalized volume density again fixes $N$:
\begin{equation}
 \dd\vol_{g_{N,\vartheta}}
 =N\dd\vol_h\dd t
 =N^{d+1}\dd\vol_{\bar h}\dd t.
 \label{eq:stationary-volume}
\end{equation}
It is independent of $\vartheta$ because the coframe change $(\dd t,\dd x)\mapsto(\dd t-\vartheta,\dd x)$ has determinant one.  Therefore Theorem~\ref{thm:static-volume} applies without change after $\bar h$ and $\vartheta$ are recovered.

\begin{remark}[Relation to global inverse problems]
Stationary boundary travel-time tomography recovers a metric up to stationary diffeomorphism under global geometric assumptions \cite{UhlmannYangZhou2021}; Randers boundary rigidity similarly carries potential-field and boundary-rigidity hypotheses \cite{Monkkonen2022}.  Proposition~\ref{prop:even-odd} is a different, infinitesimal statement: all local directions are directly addressable and the clock section is fixed.  It therefore has no geodesic coverage or boundary gauge obstruction.
\end{remark}

\section{The exact ADM indicatrix and cone--volume inverse}
\label{sec:adm}

Let a general Lorentzian metric be written in ADM form on a coordinate cylinder:
\begin{equation}
 g=-N(t,x)^2\dd t^2
 +h_{t,x}(\dd x+\beta(t,x)\dd t,\dd x+\beta(t,x)\dd t).
 \label{eq:adm-metric}
\end{equation}
Here $h$ is positive definite, $N>0$, and $\beta$ is the shift vector.  Assume the observer congruence is uniformly subcritical,
\begin{equation}
 h(\beta,\beta)\leq(1-\kappa)N^2,
 \qquad \kappa>0.
 \label{eq:subcritical}
\end{equation}

For $z=(t,x)$ and $v\in T_xM$, define $F_z(v)$ as the least positive $\tau$ such that $(\tau,v)$ is future null.  Put
\begin{equation}
 \Delta=N^2-h(\beta,\beta)>0,
 \qquad b(v)=h(\beta,v).
 \label{eq:Delta-b}
\end{equation}

\begin{theorem}[Exact ADM arrival indicatrix]
\label{thm:adm-indicatrix}
The future null indicatrix is
\begin{equation}
 F(v)=\frac{b(v)+\sqrt{b(v)^2+\Delta h(v,v)}}{\Delta}.
 \label{eq:adm-F}
\end{equation}
Its even and odd parts satisfy
\begin{align}
 \alpha(v)&=\frac{F(v)+F(-v)}2=\sqrt{a(v,v)},
 \label{eq:adm-alpha}\\
 \omega(v)&=\frac{F(v)-F(-v)}2=\frac{\beta_h^\flat(v)}{\Delta},
 \label{eq:adm-omega}\\
 a&=\frac{h}{\Delta}+\frac{\beta_h^\flat\otimes\beta_h^\flat}{\Delta^2}.
 \label{eq:adm-a}
\end{align}
Consequently
\begin{align}
 r&:=a-\omega\otimes\omega=\frac{h}{\Delta},\label{eq:r-def}\\
 B&:=r^{-1}\omega=\beta,\label{eq:B-def}\\
 \chi&:=\omega(B)=\frac{h(\beta,\beta)}{\Delta}.
 \label{eq:r-B-chi}
\end{align}
The conformal metric is
\begin{align}
 q_F&=-\dd t^2+\dd t\otimes\omega+\omega\otimes\dd t+r,
 \label{eq:qF}\\
 g&=\Delta q_F.
 \label{eq:g-from-qF}
\end{align}
\end{theorem}

\begin{proof}
The null equation for $(\tau,v)$ is
\begin{equation}
 -N^2\tau^2+h(v+\tau\beta,v+\tau\beta)=0,
 \label{eq:null-quadratic-start}
\end{equation}
which becomes
\begin{equation}
 \Delta\tau^2-2b(v)\tau-h(v,v)=0.
 \label{eq:null-quadratic}
\end{equation}
The unique positive root is Eq.~\eqref{eq:adm-F}.  Evaluating at $v$ and $-v$ yields Eqs.~\eqref{eq:adm-alpha}--\eqref{eq:adm-a}.  Subtraction of $\omega\otimes\omega$ gives $r=h/\Delta$, so $r(B,\cdot)=\omega$ implies $B=\beta$.  Expanding $\Delta q_F$ gives
\begin{equation}
 -\Delta\dd t^2+\dd t\otimes\beta_h^\flat
 +\beta_h^\flat\otimes\dd t+h,
 \label{eq:qF-expanded}
\end{equation}
which is exactly Eq.~\eqref{eq:adm-metric} after expanding the ADM square.
\end{proof}

\begin{figure*}[t]
\centering
\includegraphics[width=0.82\textwidth]{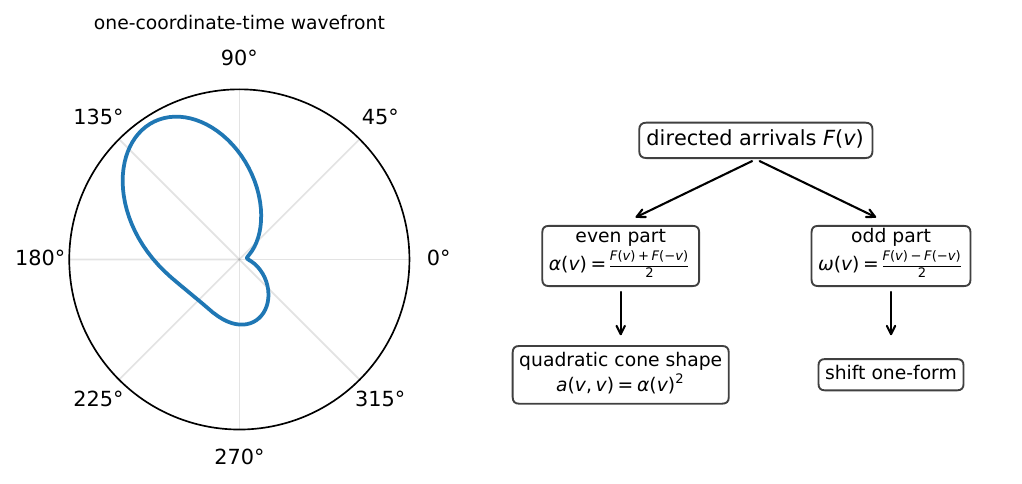}
\caption{Directed-arrival tomography.  The radius $1/F(u)$ is the spatial displacement reached in one coordinate-time unit along direction $u$.  Opposite-direction sums and differences isolate the quadratic cone shape and the shift one-form.}
\label{fig:protocol}
\end{figure*}

Figure~\ref{fig:protocol} summarizes the data flow.  The wavefront $F(u)=1$ is generally displaced and non-centrally symmetric; the even part fixes the quadratic cone shape and the odd part fixes the shift.

The local cone leaves one scalar conformal factor $\Delta$.  The volume density supplies it.

\begin{theorem}[ADM cone--volume reconstruction]
\label{thm:adm-volume}
Assume $F_z$ is known in a connected coordinate region and the normalized spacetime measure has density
\begin{equation}
 \dd\mu=\rho(t,x)\,\dd\vol_{r_t}(x)\dd t,
 \qquad \rho>0.
 \label{eq:rho-def}
\end{equation}
Then
\begin{equation}
 \rho=C_0^{-1}\Delta^{(d+1)/2}\sqrt{1+\chi},
 \label{eq:rho-Delta}
\end{equation}
so
\begin{align}
 \Delta&=C\left(\frac{\rho}{\sqrt{1+\chi}}\right)^{2/(d+1)},
 \label{eq:Delta-recovery}\\
 h&=\Delta r,\qquad \beta=B,\qquad N^2=\Delta(1+\chi).
 \label{eq:adm-recovery}
\end{align}
The constant $C>0$ is the sole ambiguity of normalized cone--volume data.
\end{theorem}

\begin{proof}
In the block decomposition $(t,x)$,
\begin{align}
 q_F&=\begin{pmatrix}-1&\omega^T\\ \omega&r\end{pmatrix},\notag\\
 \det q_F&=-(1+\omega^Tr^{-1}\omega)\det r\notag\\
 &=-(1+\chi)\det r.
 \label{eq:schur}
\end{align}
Therefore
\begin{equation}
 \dd\vol_g
 =\Delta^{(d+1)/2}\sqrt{1+\chi}\,\dd\vol_r\dd t.
 \label{eq:adm-volume-form}
\end{equation}
After normalization this gives Eq.~\eqref{eq:rho-Delta}.  Multiplying $\Delta$ by a positive constant multiplies $g$ by the same constant and leaves both $F$ and normalized volume unchanged.
\end{proof}

\begin{corollary}[Fixed-gauge identifiability modulo scale]
\label{cor:fixed-gauge-identifiability}
Let \(g\) and \(g'\) be subcritical ADM metrics on the same finite-volume
coordinate region, with the clock coordinate and spatial labels fixed.  They
have identical local directed-arrival indicatrices at every event and the
same normalized spacetime measure if and only if
\begin{equation}
 g'=c g
 \label{eq:global-scale-equivalence}
\end{equation}
for one constant \(c>0\).  It is enough to equate the \(p_d\) measurements of
Theorem~\ref{thm:minimal-inverse} at every event instead of the full
indicatrices.
\end{corollary}

\begin{proof}
Equality of the indicatrices gives the same \(a,\omega,r\), hence the same
conformal representative \(q_F\), pointwise.  Thus
\(g=\Delta q_F\) and \(g'=\Delta' q_F\).  If \(Z,Z'\) are their total
spacetime volumes, equality of the normalized measures gives
\[
 \frac{\Delta'^{(d+1)/2}}{Z'}
 =\frac{\Delta^{(d+1)/2}}{Z}.
\]
Therefore \(\Delta'/\Delta=(Z'/Z)^{2/(d+1)}=:c\) is constant and
Eq.~\eqref{eq:global-scale-equivalence} follows.  The converse is immediate
because constant conformal rescaling preserves null arrivals and cancels in
the normalized measure.  The last assertion follows from
Theorem~\ref{thm:minimal-inverse}.
\end{proof}

\begin{proposition}[Stability of the algebraic inverse]
\label{prop:adm-stability}
Fix a compact class in which $r$ is uniformly elliptic, Eq.~\eqref{eq:subcritical} holds with a common $\kappa$, and $\rho$ has common positive upper and lower bounds.  Then for sufficiently close data
\begin{equation}
 \norm{F-F'}_{C^0(S)}\leq\eta,
 \qquad
 \norm{\rho-\rho'}_{C^0}\leq\zeta,
 \label{eq:data-stability}
\end{equation}
there exists $C_*>0$ such that
\begin{equation}
 \inf_{\lambda>0}\norm{g'-\lambda g}_{C^0}
 \leq C_*(\eta+\zeta).
 \label{eq:adm-stability}
\end{equation}
If the global scale is calibrated, the infimum is removed.
\end{proposition}

\begin{proof}
The even--odd formulas give $\norm{\alpha-\alpha'}+\norm{\omega-\omega'}\leq2\eta$.  Uniform bounds on $\alpha$ imply $\norm{a-a'}\leq C\eta$ after squaring and polarizing.  The maps
\begin{equation}
 (a,\omega)\mapsto r=a-\omega\otimes\omega,
 \quad (r,\omega)\mapsto(r^{-1}\omega,\omega(r^{-1}\omega))
 \label{eq:smooth-maps}
\end{equation}
are Lipschitz on the compact elliptic class.  Equation~\eqref{eq:adm-recovery} is likewise Lipschitz after one common scale normalization.
\end{proof}

\section{A minimal finite-direction tomography}
\label{sec:minimal}

The preceding formulas appear to require the full function $F$ on the unit sphere.  Lorentzian cones are much more rigid: $F$ belongs to the finite-dimensional Randers family
\begin{equation}
 F(v)=\sqrt{a(v,v)}+\omega(v),
 \label{eq:abstract-randers}
\end{equation}
where $a$ is positive definite and $\norm{\omega}_{a^*}<1$.  In a fixed local frame $e_1,\ldots,e_d$, define the measurements
\begin{align}
 y_i^+&=F(e_i),\qquad y_i^-=F(-e_i),\label{eq:minimal-data-pm}\\
 y_{ij}&=F(e_i+e_j),\qquad i<j.
 \label{eq:minimal-data}
\end{align}
Their number is
\begin{equation}
 2d+\frac{d(d-1)}2=\frac{d(d+3)}2=p_d.
 \label{eq:measurement-count}
\end{equation}

\begin{theorem}[Exact finite-direction inverse]
\label{thm:minimal-inverse}
The data in Eqs.~\eqref{eq:minimal-data-pm} and \eqref{eq:minimal-data} recover $a$ and $\omega$ uniquely by
\begin{align}
 \omega_i&=\frac{y_i^+-y_i^-}{2},
 \label{eq:omega-components}\\
 a_{ii}&=\left(\frac{y_i^++y_i^-}{2}\right)^2,
 \label{eq:a-diagonal}\\
 a_{ij}&=\frac12\left[
 \bigl(y_{ij}-\omega_i-\omega_j\bigr)^2-a_{ii}-a_{jj}
 \right],\qquad i<j.
 \label{eq:a-offdiag}
\end{align}
For an ADM indicatrix, Eqs.~\eqref{eq:r-B-chi}--\eqref{eq:qF} then recover the conformal Lorentzian metric.
\end{theorem}

\begin{proof}
The first two formulas are the even--odd split at $e_i$.  Since $\omega(e_i+e_j)=\omega_i+\omega_j$,
\begin{equation}
 y_{ij}-\omega_i-\omega_j
 =\sqrt{a(e_i+e_j,e_i+e_j)}.
 \label{eq:pair-alpha}
\end{equation}
Squaring and subtracting the diagonal terms gives Eq.~\eqref{eq:a-offdiag}.
\end{proof}

\begin{theorem}[Exact semialgebraic validity test for finite arrivals]
\label{thm:admissible-arrival-data}
Consider arbitrary real measurements
$\{y_i^+,y_i^-,y_{ij}\}$ in the
$p_d=d(d+3)/2$ directions of
Theorem~\ref{thm:minimal-inverse}.  Define
\begin{align}
 u_i&=(y_i^++y_i^-)/2,&
 b_i&=(y_i^+-y_i^-)/2,\nonumber\\
 v_{ij}&=y_{ij}-b_i-b_j,&
 A_{ii}&=u_i^2,\nonumber\\
 A_{ij}&=(v_{ij}^2-u_i^2-u_j^2)/2,&&i<j.
 \label{eq:arrival-certificate-data}
\end{align}
These measured numbers arise from a (necessarily unique) future
Randers indicatrix $F(v)=\sqrt{v^TAv}+b^Tv$ with
$A\succ0$ and $b^TA^{-1}b<1$ if and only if
\begin{equation}
 \boxed{u_i>0,\quad v_{ij}>0,\quad
 \begin{pmatrix}1&b^T\\b&A\end{pmatrix}\succ0.}
 \label{eq:arrival-admissibility-lmi}
\end{equation}
When these inequalities hold, the data determine a unique
subcritical ADM null cone in the prescribed clock and spatial frame,
up to one positive conformal factor.  If any test fails, the data
cannot be exact measurements of such a cone; this supplies a finite,
model-falsifying consistency certificate.
\end{theorem}

\begin{proof}
Necessity of $u_i>0$ and $v_{ij}>0$ follows from
$u_i=\sqrt{A_{ii}}$ and
$v_{ij}=\sqrt{A_{ii}+A_{jj}+2A_{ij}}$.
For a Randers indicatrix, $A\succ0$ and
$b^TA^{-1}b<1$; by the Schur-complement criterion these
are equivalent to the block positivity in
\eqref{eq:arrival-admissibility-lmi}.

Conversely, suppose the inequalities hold.  The Schur complement gives
$r:=A-bb^T\succ0$, so $A\succ0$ and
$b^TA^{-1}b<1$.  Thus
$F(v)=\sqrt{v^TAv}+b^Tv$ is positive for every $v\ne0$ and
is a strictly convex Randers norm.  Evaluating at $\pm e_i$
reproduces $y_i^\pm$: the positive square root is $u_i>0$.
Similarly, evaluating at $e_i+e_j$ gives
$\sqrt{u_i^2+u_j^2+2A_{ij}}+b_i+b_j$
$=v_{ij}+b_i+b_j=y_{ij}$ because $v_{ij}>0$.
The coefficients are unique by
Eqs.~\eqref{eq:omega-components}--\eqref{eq:a-offdiag}.
Finally, choose conformal factor $\Delta=1$, spatial metric
$h=r$, shift $\beta=r^{-1}b$, and lapse
$N^2=1+b^Tr^{-1}b$.  These satisfy $N^2-h(\beta,\beta)=1>0$ and
reproduce the given indicatrix by
Theorem~\ref{thm:adm-indicatrix}.  Multiplying the Lorentzian
metric by any positive factor preserves it.
\end{proof}

\begin{remark}[The physical meaning of the matrix test]
The constraints in \eqref{eq:arrival-admissibility-lmi} are
semialgebraic in the directly observed numbers.
The block matrix is positive definite, not merely semidefinite:
a zero eigenvalue corresponds to a degenerate cone or the boundary
of the subcritical class.  A noisy dataset outside the strict feasible
region requires an explicitly calibrated projection or rejection;
substituting it blindly into the squared inverse formulas could
produce a spurious Lorentzian metric.
\end{remark}

\begin{corollary}[Certified robustness and exclusion under measurement error]
\label{cor:arrival-data-robustness}
Let $d\ge2$ and let $y$ denote one full collection of the
$p_d=d(d+3)/2$ arrival measurements.  Write $u_i$, $v_{ij}$,
$b$, $A$ and $H=\begin{psmallmatrix}1&b^T\\ b&A\end{psmallmatrix}$
for its reconstructed quantities from
\eqref{eq:arrival-certificate-data}, and put
\begin{align}
 U&=\max_i|u_i|,\qquad V=\max_{i<j}|v_{ij}|,\notag\\
 C_1(\delta)&=2U\delta+\delta^2,\notag\\
 C_2(\delta)&=(3V+2U)\delta+\tfrac{11}{2}\delta^2,\notag\\
 \mathcal E_d(\delta)&=(d+1)\max\{\delta,C_1(\delta),C_2(\delta)\}.
 \label{eq:arrival-error-envelope}
\end{align}
For any other measurement vector $y'$ whose individual entries differ
from those of $y$ by at most $\delta\ge0$, the corresponding block
matrix satisfies
\begin{equation}
 \|H(y')-H(y)\|_2\le\mathcal E_d(\delta).
 \label{eq:arrival-matrix-lipschitz}
\end{equation}
Consequently, the following explicit strict-margin conditions certify
that \emph{every} datum in the $\ell^\infty$ error ball is realized by
a unique subcritical Randers indicatrix:
\begin{equation}
 \boxed{\min_i u_i>\delta,\quad
 \min_{i<j}v_{ij}>3\delta,\quad
 \lambda_{\min}(H)>\mathcal E_d(\delta).}
 \label{eq:arrival-robust-cert}
\end{equation}
Conversely, if any one of
$\min_i u_i<-\delta$, $\min_{i<j}v_{ij}<-3\delta$, or
$\lambda_{\min}(H)<-\mathcal E_d(\delta)$ holds,
no exact subcritical Randers dataset lies in that error ball.
Failure of the sufficient inequalities does not by itself imply
incompatibility.
\end{corollary}

\begin{proof}
Since the transforms in \eqref{eq:arrival-certificate-data} are affine,
$|u_i'-u_i|,|b_i'-b_i|\le\delta$ and
$|v_{ij}'-v_{ij}|\le3\delta$.  Squaring gives
\begin{align*}
 |A_{ii}'-A_{ii}|&\le2|u_i|\delta+\delta^2,\\
 |A_{ij}'-A_{ij}|&\le\frac{6|v_{ij}|\delta+9\delta^2
 +2(|u_i|+|u_j|)\delta+2\delta^2}{2}\\
 &\le(3V+2U)\delta+\tfrac{11}{2}\delta^2.
\end{align*}
The entrywise bound and
$\|B\|_2\le\|B\|_{\mathrm F}\le(d+1)\max_{ab}|B_{ab}|$
prove \eqref{eq:arrival-matrix-lipschitz}.  Weyl's eigenvalue
inequality preserves positive definiteness under
\eqref{eq:arrival-robust-cert}; the componentwise margins preserve the
chosen signs of all square roots.  Theorem~\ref{thm:admissible-arrival-data}
then reconstructs every $y'$ in the ball.  Conversely, each displayed
strict negative-margin condition persists throughout the ball, by the
same Lipschitz or Weyl estimate, and violates the necessary conditions
of that theorem.  No converse claim is made for the intermediate zone.
\end{proof}

\begin{theorem}[Local measurement minimality]
\label{thm:minimality}
Let $\mathfrak R_d$ denote the open parameter manifold of pairs $(a,\omega)$ with $a>0$ and $\norm{\omega}_{a^*}<1$.  It has dimension $p_d=d(d+3)/2$.  A fixed-direction protocol consisting of $m$ scalar evaluations
\begin{equation}
 (a,\omega)\longmapsto\bigl(F(v_1),\ldots,F(v_m)\bigr)
 \label{eq:evaluation-map}
\end{equation}
that admits a $C^1$ local inverse on an open subset of $\mathfrak R_d$ must satisfy $m\geq p_d$.  The protocol of Theorem~\ref{thm:minimal-inverse} is therefore $C^1$-locally minimal.
\end{theorem}

\begin{proof}
The parameter count follows from $d(d+1)/2$ symmetric coefficients of $a$ and $d$ coefficients of $\omega$.  If the evaluation map had a $C^1$ local inverse, its differential would have rank $p_d$.  But a map into $\RR^m$ has differential rank at most $m$, so $m\geq p_d$.
\end{proof}

\begin{proposition}[Noise stability of the minimal inverse]
\label{prop:minimal-noise}
On any compact subset of $\mathfrak R_d$, there are $\epsilon_0,C>0$ such that if every datum in Eqs.~\eqref{eq:minimal-data-pm} and \eqref{eq:minimal-data} is perturbed by at most $\epsilon<\epsilon_0$, then the reconstructed coefficients satisfy
\begin{equation}
 \norm{\widehat a-a}+\norm{\widehat\omega-\omega}
 +\norm{\widehat q_F-q_F}\leq C\epsilon.
 \label{eq:minimal-noise}
\end{equation}
\end{proposition}

\begin{proof}
Equations~\eqref{eq:omega-components}--\eqref{eq:a-offdiag} are polynomial expressions in the measurements after one subtraction.  On a compact class all measured values are uniformly bounded, so the coefficient map is Lipschitz.  For small enough $\epsilon$, $\widehat r=\widehat a-\widehat\omega\otimes\widehat\omega$ remains uniformly positive, and inversion of $r$ is Lipschitz.
\end{proof}

In $d=3$, nine one-way measurements recover the nine conformal ADM coefficients in a fixed spatial frame.  The volume density adds the single conformal scale field, modulo its global normalization.

\section{Short-time tomography in nonstationary geometry}
\label{sec:short-time}

The local algebra assumes $F_z$ is known.  We now derive it from finite displacements.  Let $U\subset\RR^d$ be a convex coordinate neighborhood and suppose
\begin{equation}
 F:(t,x,v)\mapsto F_{t,x}(v)
 \label{eq:F-family}
\end{equation}
obeys the following uniform conditions on a compact subcylinder:
\begin{align}
 m\abs v&\leq F_z(v)\leq M\abs v,
 \label{eq:F-equivalence}\\
 \abs{F_z(v)-F_{z'}(v)}&\leq L\abs{z-z'}\abs v,
 \label{eq:F-Lipschitz}
\end{align}
with $m,M>0$, and each $F_z$ is convex and positively homogeneous.  A curve parametrized by coordinate time is causal exactly when
\begin{equation}
 F_{t+s,x(s)}(\dot x(s))\leq1.
 \label{eq:control-constraint}
\end{equation}

\begin{theorem}[Short-time freezing]
\label{thm:freezing}
For every unit vector $v$ and sufficiently small $\epsilon>0$,
\begin{equation}
 \abs{\cT_z(\epsilon v)-\epsilon F_z(v)}\leq C\epsilon^2,
 \label{eq:freezing}
\end{equation}
uniformly on compact subsets.  The constant depends only on $m,M,L$ and the chart margin.
\end{theorem}

\begin{proof}
Let $\tau=\cT_z(\epsilon v)+o(1)$ and let $x(s)$ be an almost minimizing causal path for $0\leq s\leq\tau$.  Equation~\eqref{eq:F-equivalence} gives $\abs{\dot x}\leq m^{-1}$ and therefore $\abs{x(s)-x}=O(\tau)$.  An explicit straight competitor below shows $\tau=O(\epsilon)$, so Eq.~\eqref{eq:F-Lipschitz} yields
\begin{equation}
 F_z(\dot x(s))\leq1+C\epsilon\abs{\dot x(s)}.
 \label{eq:frozen-bound}
\end{equation}
By convexity and homogeneity,
\begin{equation}
 \epsilon F_z(v)
 =F_z\left(\int_0^\tau\dot x(s)\dd s\right)
 \leq\int_0^\tau F_z(\dot x(s))\dd s
 \leq\tau+C\epsilon\tau,
 \label{eq:freezing-lower}
\end{equation}
which gives $\tau\geq\epsilon F_z(v)-C\epsilon^2$.

For the upper bound, take the straight spatial path $x(s)=x+s\epsilon v/\tau$ and set
\begin{equation}
 \tau=\epsilon F_z(v)+K\epsilon^2.
 \label{eq:straight-time}
\end{equation}
Along it, $z_s-z=O(\epsilon)$ and
\begin{equation}
 F_{z_s}\left(\frac{\epsilon v}{\tau}\right)
 \leq\frac{\epsilon(F_z(v)+C\epsilon)}{\epsilon F_z(v)+K\epsilon^2}
 \leq1
 \label{eq:straight-causal}
\end{equation}
for a uniform $K$.  Hence the path is causal and gives the matching upper bound.
\end{proof}

\begin{figure*}[t]
\centering
\includegraphics[width=0.62\textwidth]{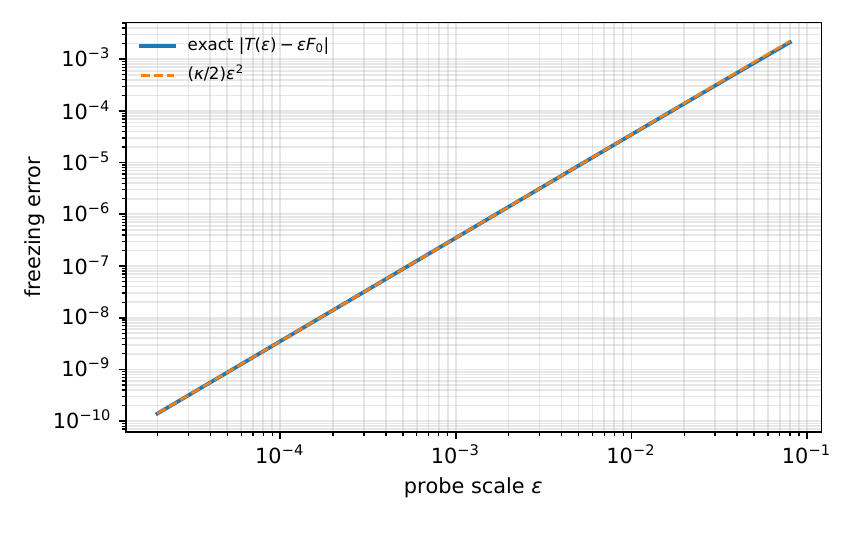}
\caption{Exact short-time arrival error for Eq.~\eqref{eq:1d-example} with $\kappa=0.7$.  The fitted log--log slope is $1.9988$ and the scaled coefficient tends to $\kappa/2=0.35$.}
\label{fig:freezing}
\end{figure*}

The quadratic rate is sharp already in one dimension.  For
\begin{equation}
 g=-(1+\kappa t)^2\dd t^2+\dd x^2,
 \label{eq:1d-example}
\end{equation}
light emitted at $(0,0)$ reaches $x=\epsilon>0$ at
\begin{equation}
 \cT(\epsilon)=\frac{\sqrt{1+2\kappa\epsilon}-1}{\kappa}
 =\epsilon-\frac{\kappa}{2}\epsilon^2+O(\epsilon^3).
 \label{eq:1d-arrival}
\end{equation}
Figure~\ref{fig:freezing} verifies this scaling over four decades.

\section{Finite-resolution ADM reconstruction}
\label{sec:finite-resolution}

At an event $z$, use the minimal directions
\begin{equation}
 \cD=\{\pm e_i:1\leq i\leq d\}
 \cup\{e_i+e_j:1\leq i<j\leq d\}.
 \label{eq:direction-set}
\end{equation}
For a probe scale $\epsilon$, define normalized measurements
\begin{equation}
 \widehat y(v)=\frac{\widehat\cT_z(\epsilon v)}{\epsilon}.
 \label{eq:normalized-measurement}
\end{equation}
Theorem~\ref{thm:freezing} and Eq.~\eqref{eq:oracle-error} give
\begin{equation}
 \abs{\widehat y(v)-F_z(v)}
 \leq\frac{a}{\epsilon}+C\epsilon,
 \qquad v\in\cD.
 \label{eq:normalized-error}
\end{equation}

\begin{theorem}[Minimal finite-resolution ADM tomography]
\label{thm:finite-resolution}
Fix a compact subcritical ADM class as in Proposition~\ref{prop:adm-stability}.  Suppose the $p_d$ arrivals in Eq.~\eqref{eq:direction-set} satisfy Eq.~\eqref{eq:oracle-error}, and a density estimator obeys
\begin{equation}
 \norm{\widehat\rho-\rho}_{C^0}\leq\zeta.
 \label{eq:density-error}
\end{equation}
Then the metric reconstructed by Eqs.~\eqref{eq:omega-components}--\eqref{eq:a-offdiag}, \eqref{eq:r-def}--\eqref{eq:r-B-chi}, and \eqref{eq:Delta-recovery}--\eqref{eq:adm-recovery} satisfies
\begin{equation}
 \inf_{\lambda>0}\norm{\widehat g-\lambda g}_{C^0}
 \leq C_*
 \left(\frac{a}{\epsilon}+\epsilon+\zeta\right).
 \label{eq:metric-rate}
\end{equation}
If the overall scale is calibrated, the infimum is removed.  Choosing
\begin{equation}
 \epsilon\asymp\sqrt a
 \label{eq:optimal-epsilon}
\end{equation}
gives
\begin{equation}
 \inf_{\lambda>0}\norm{\widehat g-\lambda g}_{C^0}
 =O(\sqrt a+\zeta).
 \label{eq:sqrt-rate}
\end{equation}
\end{theorem}

\begin{proof}
Equation~\eqref{eq:normalized-error} provides a uniform perturbation of the minimal data.  Proposition~\ref{prop:minimal-noise} converts it into an $O(a/\epsilon+\epsilon)$ error for the conformal metric.  Proposition~\ref{prop:adm-stability} adds the density error.  The first two terms are minimized at the square-root scale.
\end{proof}

This theorem improves the generic directional-net estimate.  If no Lorentzian/Randers model is assumed and $F$ is only a smooth convex indicatrix, a $\delta$-net and interpolation produce the additional error $O(\delta)$.  The Lorentzian quadratic-plus-linear structure collapses the continuum sphere to the exact finite set $\cD$.

\begin{figure*}[t]
\centering
\includegraphics[width=0.465\textwidth]{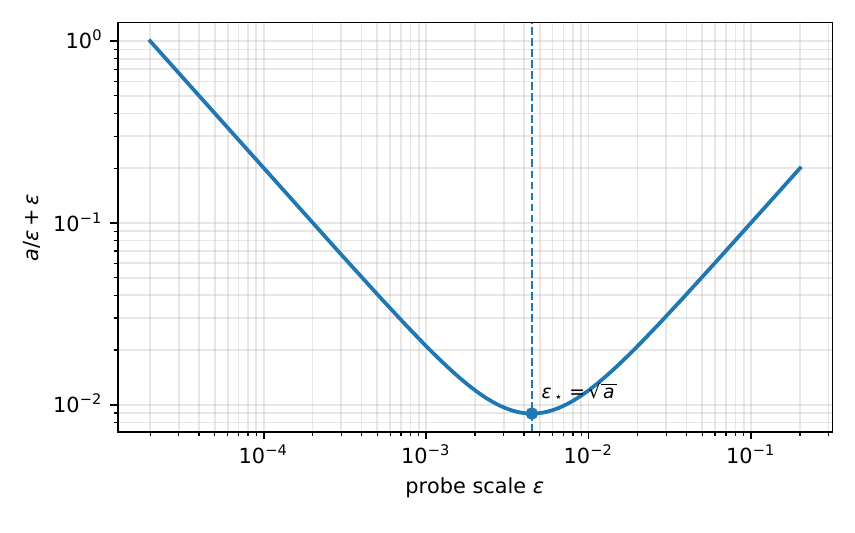}\hfill
\includegraphics[width=0.465\textwidth]{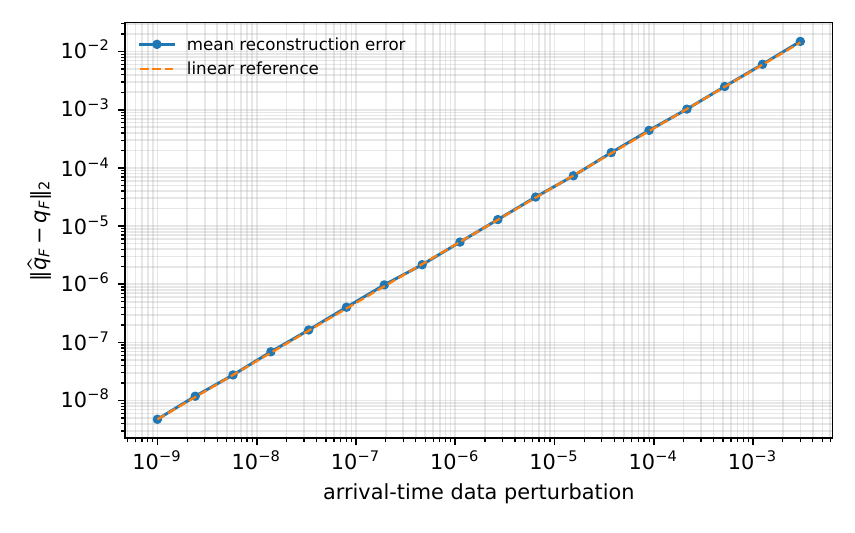}
\caption{Finite-resolution behavior.  (a) Bias--noise competition: absolute timing error contributes $a/\epsilon$, while freezing the time-dependent cone contributes $O(\epsilon)$, with optimum $\epsilon_\star=\sqrt a$.  (b) Numerical perturbation test of the minimal $d=3$ inverse: nine directed measurements reconstruct the conformal ADM metric with fitted small-noise exponent $0.99944$.}
\label{fig:finite-resolution-audit}
\end{figure*}

\section{From arrival errors to causal-order errors}
\label{sec:order-error}

A finite arrival-time estimate induces a binary causal relation.  The only pairs whose indicator can change are those lying close to the null boundary.  The following elementary anti-concentration statement makes this precise.

\begin{lemma}[Conditional temporal anti-concentration]
\label{lem:anticoncentration}
Let \(Z\) denote arbitrary spatial data and suppose the conditional law of
\(U=T-S\) given \(Z\) has a density \(g_Z\) satisfying
\(\|g_Z\|_\infty\le M_t\) almost surely.  Then for every measurable centre
\(u_0(Z)\) and every \(\delta>0\),
\begin{equation}
 \Pr\{\abs{U-u_0(Z)}\leq\delta\}
 \leq2M_t\delta.
\label{eq:anticoncentration}
\end{equation}
In particular, the hypothesis holds when \(S,T\) are independent of \(Z\),
independent of each other, and have a common density bounded by \(M_t\).
\end{lemma}

\begin{proof}
Condition on \(Z\) and integrate \(g_Z\) over an interval of length
\(2\delta\).  In the stated independent special case the difference density
is \(g(u)=\int f(s)f(s+u)\dd s\le M_t\).
\end{proof}

\begin{proposition}[Stationary causal-pair mismatch]
\label{prop:pair-mismatch}
Suppose the conditional density hypothesis of
Lemma~\ref{lem:anticoncentration} holds with
\(Z=(X,Y,\widehat d_F(X,Y))\), and a directed distance
estimate $\widehat d_F(X,Y)$ satisfies
\begin{equation}
 \abs{\widehat d_F(X,Y)-d_F(X,Y)}\leq\delta
 \label{eq:distance-error}
\end{equation}
outside an event of probability $e$.  Let the true and estimated chronological indicators be
\begin{align}
 C&=\one\{d_F(X,Y)<T-S\},
 \label{eq:true-indicator}\\
 \widehat C&=\one\{\widehat d_F(X,Y)<T-S\}.
 \label{eq:estimated-indicator}
\end{align}
Then
\begin{equation}
 \Pr\{C\neq\widehat C\}\leq e+2M_t\delta.
 \label{eq:pair-mismatch}
\end{equation}
For uniform clock readings on $[0,T_0]$, $M_t=1/T_0$.
\end{proposition}

\begin{proof}
Let \(E\) be the event on which Eq.~\eqref{eq:distance-error} holds.  On
\(E\), a disagreement can occur only when \(U=T-S\) lies between
\(d_F(X,Y)\) and \(\widehat d_F(X,Y)\).  Hence
\[
 \{C\ne\widehat C\}\cap E
 \subseteq
 \{\abs{U-d_F(X,Y)}\le\delta\}.
\]
Condition on \((X,Y,\widehat d_F)\), apply
Lemma~\ref{lem:anticoncentration}, and then add \(\Pr(E^c)\le e\).
\end{proof}

This is the correct interface between local metric error and a causal-profile statistic.  It does not imply global geometric uniqueness without a compact admissible class and an inverse modulus; those statistical and minimax issues are treated separately in Ref.~\cite{ChurilovFiniteResolution2026}.

\section{Deterministic audit}
\label{sec:audit}

The source package contains \texttt{anc/verify\_optical\_adm.py}.  It performs five independent checks.

\begin{enumerate}[label=(\arabic*)]
\item A random three-dimensional positive tensor $h$, subcritical shift $\beta$, and lapse $N$ are generated with fixed seed.  Nine values in the minimal direction set reconstruct $a$, $\omega$, $r$, $\beta$, $\chi$, and $q_F$.
\item The Schur determinant identity
\begin{equation}
 \det q_F=-(1+\chi)\det r
 \label{eq:audit-det}
\end{equation}
is checked directly.
\item The exactly solvable nonstationary example \eqref{eq:1d-example} is fitted on a log--log grid to test the quadratic freezing law.
\item Uniform perturbations are added to all nine arrival data and the reconstructed matrix error is fitted against noise amplitude.
\item A constant conformal rescaling is verified to leave normalized volume
      unchanged, while a seeded nonconstant rescaling is detected.
\end{enumerate}

The maximum algebraic residual is $1.04\times10^{-15}$, the determinant residual is $5.33\times10^{-15}$, the freezing exponent is $1.99880$, and the noise exponent is $0.99944$.  The conformal negative control distinguishes constant from nonconstant rescaling.  The machine-readable record is \texttt{anc/verification\_summary.json}.  These computations validate the formulas and scaling examples; they are not substitutes for the proofs.

\section{Discussion and limits}
\label{sec:discussion}

The reconstruction separates three layers that are often conflated.

\emph{Cone layer.}  Directed short-range arrivals determine the future null indicatrix.  In a Lorentzian ADM class, the indicatrix is not an arbitrary convex function: it is a quadratic norm plus a linear one-form.  This rigidity is what permits the minimal finite measurement theorem.

\emph{Scale layer.}  The cone determines only $[g]$.  A normalized volume density determines the conformal factor pointwise but retains one global constant, exactly as expected from the invariance of normalized order--volume data under $g\mapsto Cg$.

\emph{Dynamical layer.}  The geometric inverse does not by itself prove that a microscopic quantum model supplies the required arrival oracle.  The two-sided flat and finite-volume propagation theorems of Refs.~\cite{ChurilovSharpCones2026,ChurilovFiniteVolume2026} provide controlled examples, but a random curved phase requires a uniform inner-cone theorem, clock synchronization, and composability of local measurements.

Several limitations are structural rather than technical.

First, the shift is observer and slicing dependent.  The paper reconstructs it in the declared clock gauge; no invariant meaning is assigned to $\beta$ alone.  Second, the local protocol assumes spatial displacements $e_i$ can be addressed and compared.  In a fully relational theory, constructing that frame is a separate descent problem.  Third, the short-time theorem is a $C^0$ metric reconstruction.  Curvature requires differentiated data and correspondingly stronger regularity and noise control.  Fourth, the volume density must be estimated relative to the reconstructed $r$-volume; the theorem does not prescribe a universal finite-sample estimator.

Within these boundaries, the result is complete: a general time-dependent Lorentzian metric in a fixed local observer gauge can be reconstructed, up to its unavoidable global scale, from a minimal finite set of directed causal arrivals and a local volume density, with an explicit optimal finite-resolution law.

\section{Conclusion}
\label{sec:conclusion}

Directed arrival times are a concrete operational coordinate on Lorentzian causal geometry.  Static arrivals recover the optical metric; stationary asymmetry recovers the clock-gauge shift; and a time-dependent ADM cone is inverted by the same even--odd algebra.  In $d$ spatial dimensions, $d(d+3)/2$ directed measurements are sufficient and locally minimal. The reconstructed data also admit an exact feasibility criterion and explicit error-ball certificates separating robustly physical data from datasets incompatible with any subcritical cone.  The nonstationary freezing error is quadratic in the probe length, while absolute timing error is amplified inversely with that length, producing the optimal square-root resolution scale.  Cone data and normalized volume then reconstruct the full metric up to one global constant.

The remaining frontier is not local algebraic tomography.  It is dynamical: deriving a uniformly accurate, directionally addressable arrival oracle and compatible clock foliation from one background-free quantum phase.

\paragraph*{Computational and language-model assistance.}
Generative-AI language tools assisted with manuscript editing and selected
symbolic and numerical verification scripts. These checks certify only the
explicitly stated computational tests and do not independently validate
all proofs. The author is responsible for the scientific statements,
references, and conclusions.

\appendix

\section{Coordinate covariance of the finite-direction formulas}
\label{app:covariance}

The explicit protocol uses a local frame because measurements must be assigned directions.  Fix the convention that the old and new component columns satisfy $v=A v'$.  Then the reconstructed coefficients transform as
\begin{equation}
 a'=A^TaA,\qquad \omega'=A^T\omega.
 \label{eq:frame-transform}
\end{equation}
This is exactly the tensorial pullback law for a bilinear form and a covector.  The tensor $r=a-\omega\otimes\omega$ and the Lorentzian form $q_F$ are therefore independent of the chosen computational basis.  What is physical is the calibrated set of tangent vectors used by the probes, not the coordinate values $e_i$ themselves.

\section{Why the one-sided pair directions are sufficient}
\label{app:one-sided}

A symmetric quadratic form ordinarily requires evaluations on $e_i$ and $e_i+e_j$.  The nonreversible term seems to double the data, but opposite measurements are needed only on the basis vectors.  Once $\omega_i$ are known from $F(\pm e_i)$, the odd contribution to $F(e_i+e_j)$ is already known.  Therefore only one orientation of each pair direction is required.  This is what reduces
\begin{equation}
 2d+d(d-1)
 \label{eq:naive-count}
\end{equation}
to the parameter-optimal count $2d+d(d-1)/2$.

\section{Explicit perturbation bounds for the algebraic protocol}
\label{app:explicit-noise}

Suppose $\abs{\widehat y-y}\leq\epsilon$ for every datum, and let $Y$ bound all unperturbed and perturbed data.  Then
\begin{equation}
 \abs{\widehat\omega_i-\omega_i}\leq\epsilon,
 \qquad
 \abs{\widehat a_{ii}-a_{ii}}\leq2Y\epsilon.
 \label{eq:diag-noise}
\end{equation}
For $i<j$, the inferred even value
\begin{equation}
 \widehat\alpha_{ij}=\widehat y_{ij}-\widehat\omega_i-\widehat\omega_j
 \label{eq:alphaij-hat}
\end{equation}
has error at most $3\epsilon$.  Moreover, $|\alpha_{ij}|$ and $|\widehat\alpha_{ij}|$ are at most $3Y$, and therefore
\begin{equation}
 \abs{\widehat a_{ij}-a_{ij}}
 \leq 11Y\epsilon.
 \label{eq:offdiag-noise}
\end{equation}  Thus $\norm{\widehat a-a}\leq C_dY\epsilon$.  If $r\geq r_-\one$, then for $\epsilon$ small enough $\widehat r\geq r_-\one/2$ and
\begin{equation}
 \norm{\widehat r^{-1}-r^{-1}}
 \leq\frac{2}{r_-^2}\norm{\widehat r-r}.
 \label{eq:inverse-noise}
\end{equation}
These elementary estimates supply explicit constants for Proposition~\ref{prop:minimal-noise}.

\section{Reproducibility files}
\label{app:reproducibility}

The arXiv package contains:
\begin{itemize}
\item \texttt{main.tex}, \texttt{references.bib}, and a frozen \texttt{main.bbl};
\item vector PDF figures in \texttt{figures/};
\item \texttt{anc/verify\_optical\_adm.py};
\item \texttt{anc/verification\_summary.json};
\item environment notes and a dependency list.
\end{itemize}
The audit is deterministic and requires only NumPy and Matplotlib.

\bibliographystyle{apsrev4-2}
\bibliography{references}

\end{document}